\documentclass[aps,prl,floatfix,showpacs,amsmath,amssymb,twocolumn]{revtex4-1}
\usepackage{latexsym}
\usepackage{amsfonts}
\usepackage{amssymb}
\usepackage{amsbsy}
\usepackage{newtxtext,newtxmath,amsmath}
\usepackage{dsfont}
\usepackage{graphicx}
\usepackage{color}
\usepackage{float}
\usepackage{graphicx}
\usepackage{bm}
\usepackage{natbib}

\begin{document}

\title{Diffusion of Fractal Particles in a Fractal Fluid}
\author{Marco Heinen}
\email[]{marco@marco-heinen.de}
\affiliation{
Independent researcher at the time of submission.\\
Most of the reported work was conducted at the\\~\\
Departamento de Ingenier\'{i}a F\'{i}sica,
Divisi\'{o}n de Ciencias e Ingenier\'{i}as,
University of Guanajuato,
Loma del Bosque 103,
37150 Le\'{o}n,
M\'{e}xico
}

\date{\today}

\begin{abstract}
Anomalous short- and long-time self-diffusion of non-overlapping fractal particles on a percolation cluster with spreading dimension $1.67(2)$
is studied by dynamic Monte Carlo simulations. As reported in Phys. Rev. Lett. \textbf{115}, 097801 (2015),
the disordered phase formed by these particles is that of an unconfined, homogeneous and monodisperse fluid in fractal space.
During particle diffusion in thermodynamic equilibrium, the mean squared chemical displacement increases as a nonlinear power of time,
with an exponent of $0.96(1)$ at short times and $0.63(1)$ at long times.
At finite packing fractions the steric hindrance among nearest neighbor particles leads to a sub-diffusive regime
that separates short-time anomalous diffusion from long-time anomalous diffusion.
Particle localization is observed over eight decades in time for packing fractions of $\sim 60\%$ and higher. 
\end{abstract}

\maketitle

Brownian particles exist in various spatial dimensions $d$. In the standard case $d=3$, particles are free to move in a container
with length scales that greatly exceed the particle size as well as the average step length of the random walk trajectories. Particles that are
confined in narrow slits \cite{Neser1997, Steward2014} or at interfaces \cite{Zahn2000, Deutschlaender2013} experience diffusion in $d=2$,
and a further confinement to channels \cite{Hahn1996, Wei2000, Lutz2004} gives rise to the peculiar single-file diffusion in $d=1$. 
More generally, the configuration-space dimension of Brownian particles need not be an integer number: Diffusion of particles in fractal-dimensional
porous media is a central problem of percolation theory \cite{benAvraham_Havlin2000} and the topic of the present letter. 
The qualitative characteristics of Brownian motion depend both on the spatial dimension and the existence of an intrinsic particle-interaction length scale.

In integer dimensions $d\geq2$, non-interacting point-like particles as well as interacting,
spatially extended particles exhibit normal diffusion with a mean-squared displacement
$\left\langle\right. \delta r^2 (t) \left.\right\rangle = \left\langle\right. [ r(t) - r(0) ]^2 \left.\right\rangle = 2d ~ D(t) ~ t$ \cite{benAvraham_Havlin2000}.
Here, $\delta r$ is the Euclidean distance between the initial and terminal points $r(0)$ and $r(t)$ of a Brownian particle trajectory,
the brackets $\left\langle\right. \cdot \left.\right\rangle$ indicate an ensemble average and $t$ is the correlation time.
For non-interacting point-like particles, the self-diffusion coefficient $D(t) \equiv d_s$ is a time-independent constant.
For interacting particles with a non-zero diameter (\textit{i.e.} interaction length-scale) one has to distinguish between short-time and long-time normal diffusion
characterized by the coefficients $D_s = D(t \to 0)$ and $D_l = D(t \to \infty)$, respectively. Both $D_s$ and $D_l$ (which is smaller than $D_s$)
depend on the particle interactions and the particle concentration.

\begin{figure}
\includegraphics[width=.99\columnwidth]{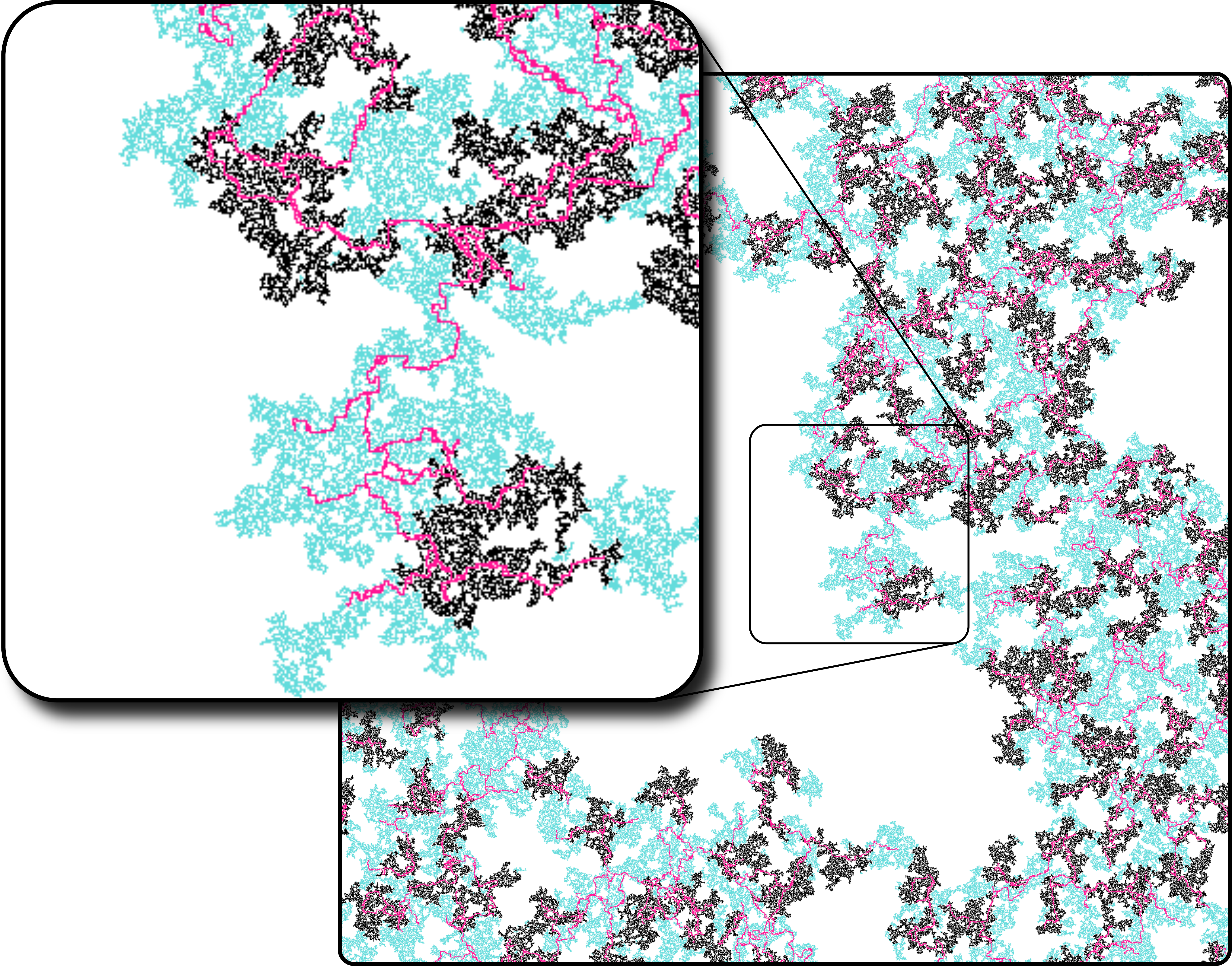}
\vspace{-0.5em}
\caption{Snapshot of a simulation box containing 112 non-overlapping fractal particles (black) on a percolation cluster (light blue)
at packing fraction $\phi=0.5$.
Every colored pixel corresponds to one vertex of the percolating configuration space.
Allowed particle center vertices are displayed in pink. The magnifying inset contains
three complete fractal particles and parts of four more particles (reaching into the inset at the upper and right boundaries).
}
\label{fig:Snapshot}
\end{figure}

Non-interacting particles in $d=1$ show normal diffusion with a time-independent self-diffusion constant if they are
allowed to interchange their positional ordering by passing through each other. In contrast, however, the interacting, distinguishable particles
of a one-dimensional single file exhibit normal diffusion at short times and anomalous diffusion at long times,
with $\left\langle\right. \delta r^2 (t \to \infty) \left.\right\rangle = F ~ \sqrt{t}$ \cite{Levitt1973, Karger1992, Wei2000, Lutz2004}.

A common model of fractal-dimensional confinement is the restriction of particle coordinates to a percolation cluster with a porosity close to the critical value
that marks the percolation transition. Such confinement is a model for naturally occurring or synthetic complex fluids and fluid mixtures in random porous media
\cite{Pape1999, Hoefling2013}. In an embedding $d=2$ Euclidean space, the critically percolating cluster shows a fractal dimension of $d_f = 91/48$
and a spreading dimension of $d_l = d_f / d_{min} = 1.67659(3)$ \cite{Zhou2012}. Here, $d_{min} = 1.13077(2)$ is the dimension of the average shortest (chemical)
path $l$ that connects two sites of the cluster with Euclidean distance $r$. Non-interacting point-like particles confined to a critical percolation cluster in an embedding $d=2$
space exhibit anomalous diffusion with $\left\langle\right. \delta r^2 (t) \left.\right\rangle \propto t^{2/d_w} = t^{0.6949(2)}$ and with a 
mean-squared chemical displacement (MSCD) of $\left\langle\right. \delta l^2 (t) \left.\right\rangle \propto t^{2/d_w^l} = t^{0.7858(3)}$ \cite{benAvraham_Havlin2000}.
Here, $d_w = 2.878(1)$ and $d_w^l = d_w/d_{min} = 2.5451(9)$ denote the walk dimensions in Euclidean and chemical space, respectively.

In this work, I report my findings on the diffusion of interacting non-overlapping fractal particles with a chemical hard-core diameter $\sigma = 200a$
on a near-critical fractal percolation cluster with lattice constant $a$.
After deleting $915.000$ random sites from an initial simple square lattice of $1500 \times 1500$ vertices, only the largest (and percolating) cluster is retained
as the particle configuration space for the simulation. All smaller clusters are deleted from the lattice. To minimize artifacts related to off-criticality, 
the spreading dimension of the percolation cluster is measured, and the cluster is accepted as a particle configuration space only if the deviation from
$d_l = 1.67659$ is less than $1\%$, with a $99\%$ confidence interval. All measured exponents are therefore reported with a $1\%$ error margin in the following.
Euclidean periodic boundary conditions are applied on the boundary of the simulation box.
Unlike confined particles that interact along Euclidean space \cite{Krakoviack2005, Skinner2013}, the present fractal particles interact via a no overlap condition that forbids
\emph{chemical} distances $l \leq \sigma$ between any two particles. Chemical distance is an intrinsic measure on the fractal percolation cluster, \textit{i.e.} a geodesic
distance that can be measured without leaving the cluster.
No reference will be made to any extrinsic measure (such as Euclidean distance) in the remaining part of this letter.

\begin{figure}
\includegraphics[width=.9\columnwidth]{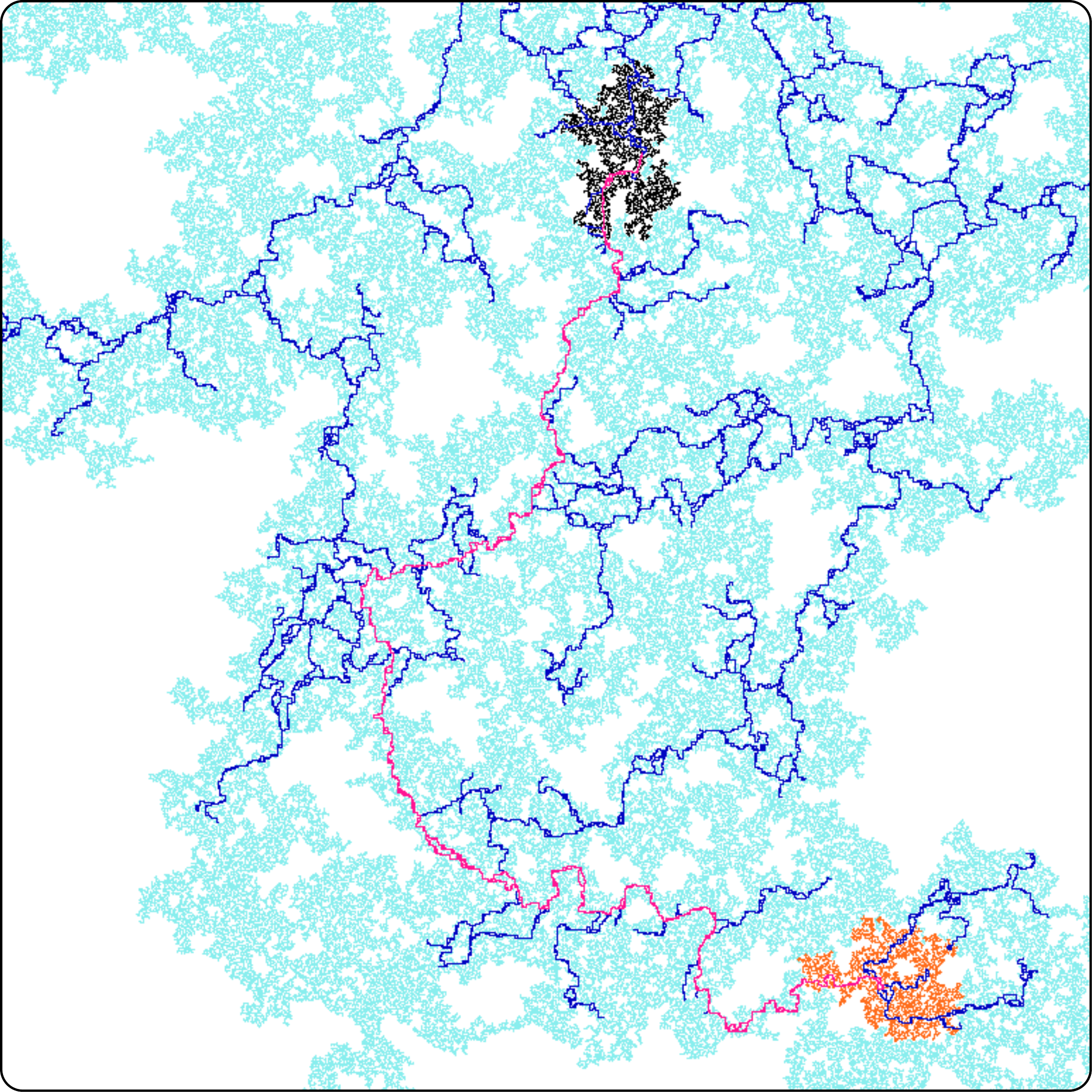}
\vspace{-0.5em}
\caption{Diffusion of a single particle from an initial position (orange) to a final position (black).
Light Blue: Percolation cluster.
Deep Blue: Particle center vertex trajectory.
Pink: Shortest (chemical) path between the initial and terminal vertices of the trajectory.
}
\label{fig:SingleParticleDiffu}
\end{figure}

In thermodynamic equilibrium, the fractal particles form a disordered fluid phase which has been characterized in Ref. \cite{Heinen2015} as a generalization of the
monodisperse, isotropic and homogeneous hard-sphere fluid from integer dimension to $d = d_l$. A representative snapshot of such a system is shown in Fig. \ref{fig:Snapshot}.
Like monodisperse hard-sphere fluids in integer dimensions, the fluid of non-overlapping fractal particles in fractal space is entirely characterized by the dimensionless
particle packing fraction $\phi$. The interested reader is referred to Ref. \cite{Heinen2015} for an explanation on how to measure the value of $\phi$ for the fractal fluid in a
Monte Carlo (MC) simulation. The pair correlations of the fractal particles are well described by the numerically solved Percus-Yevick (PY) integral equation for \emph{unconfined}
particles \cite{Percus1958}, which is analytically continued with respect to dimension \cite{Heinen2014, Heinen2015} and evaluated for $d = d_l$. Using the value of $\phi$ that is
measured in the simulation as input for the PY equation, the latter is free of any adjustable parameters. Furthermore, it represents the \emph{unique} analytic continuation
of the PY equation solution, which reduces to the numerically exact solution in all positive integer dimensions.
An alternative, analytically tractable fractal-dimensional extension of the PY equation, which reduces to the exact solution for $d \in \{1,2,3\}$ only,
has been shown to be highly accurate albeit only an approximation of the precise analytical continuation with respect to dimension \cite{Santos2016}.
To date, nothing has been known about the diffusive dynamics of the fractal particles in the fractal fluid.

As in Ref. \cite{Heinen2015}, a fractal particle is defined here as the set of vertices that belong to the percolation cluster, and that have a chemical distance
$l < \sigma/2$ from the particle center vertex. In Fig. \ref{fig:Snapshot}, every black patch represents one of $112$ fractal particles in the simulation box. 
Not all vertices of the percolation cluster qualify as a particle center vertices, since every particle is required to include at least two perimeter vertices with a mutual
chemical distance of $l=\sigma$. Pink pixels in Fig. \ref{fig:Snapshot} represent those vertices that qualify as particle centers.

The number of allowed particle-center vertices is less than $10^5$, which is approximately a factor of $10$ less than the total number of vertices in the percolation cluster.
This relatively small number of particle-center vertices renders the computation and storage of a complete chemical distances lookup table for any possible pair of particles feasible.
For the selected grid size, such a lookup table occupies less than $10$ Gigabyte, and it can be generated during less than one day of CPU time on a single core.
The lookup table entirely eliminates the need for costly chemical distance calculations \cite{Dijkstra1959} during the
MC sampling of the particle ensemble in thermodynamic equilibrium. Determination of the chemical distance between particle centers is now reduced to a memory access,
rendering the simulation with lookup table orders of magnitude faster than the earlier MC simulations reported in Ref. \cite{Heinen2015}, where chemical distances have been calculated
dynamically.

The second modification of the MC simulation method from Ref. \cite{Heinen2015} consists in the replacement of global particle moves by local moves. In the new simulation
algorithm applied here, a particle move proceeds as follows: First, a particle is randomly selected and removed from the system. The center vertex of the removed particle is then
displaced in a random walk along the set of allowed particle-center vertices, with a step length of $a$ and a number of steps $n$ that is randomly picked from the interval
of integers $[1,2,\ldots8]$. If the displaced particle-center vertex shows a chemical distance of more than $\sigma$ from all other particle center vertices, then the move is accepted.
Otherwise, the move is rejected and the particle is recreated at its original position.

After an initial equilibration phase, both the MSCD and the mean rectified path (MRP) are recorded. The MRP is the sum over displacements $n$ in all accepted moves,
divided by the constant number of particles. Figure \ref{fig:SingleParticleDiffu} illustrates a representative trajectory of a single fractal particle. The non self-avoiding
particle-center trajectory (whose contour length is the MRP) traces out the deep blue pixels, while the shortest (chemical) path between the initial and terminal points of the
trajectory is outlined by the pink pixels in Fig. \ref{fig:SingleParticleDiffu}. Monte Carlo simulations with infinitesimally short local moves are equivalent to
Brownian Dynamics (BD) \cite{Cichocki1990}, and the equivalence carries over approximately for finite local step lengths that do not considerably exceed $\sim1\%$
of the particle diameter \cite{Sanz2010}. In such 'dynamic MC' simulations, also known as 'smart MC', the MRP has been shown to be proportional to the time
variable in the equivalent BD simulation \cite{Sanz2010}. It is this type of 'dynamic' or 'smart' MC which is applied here.

\begin{figure}
\includegraphics[width=.99\columnwidth]{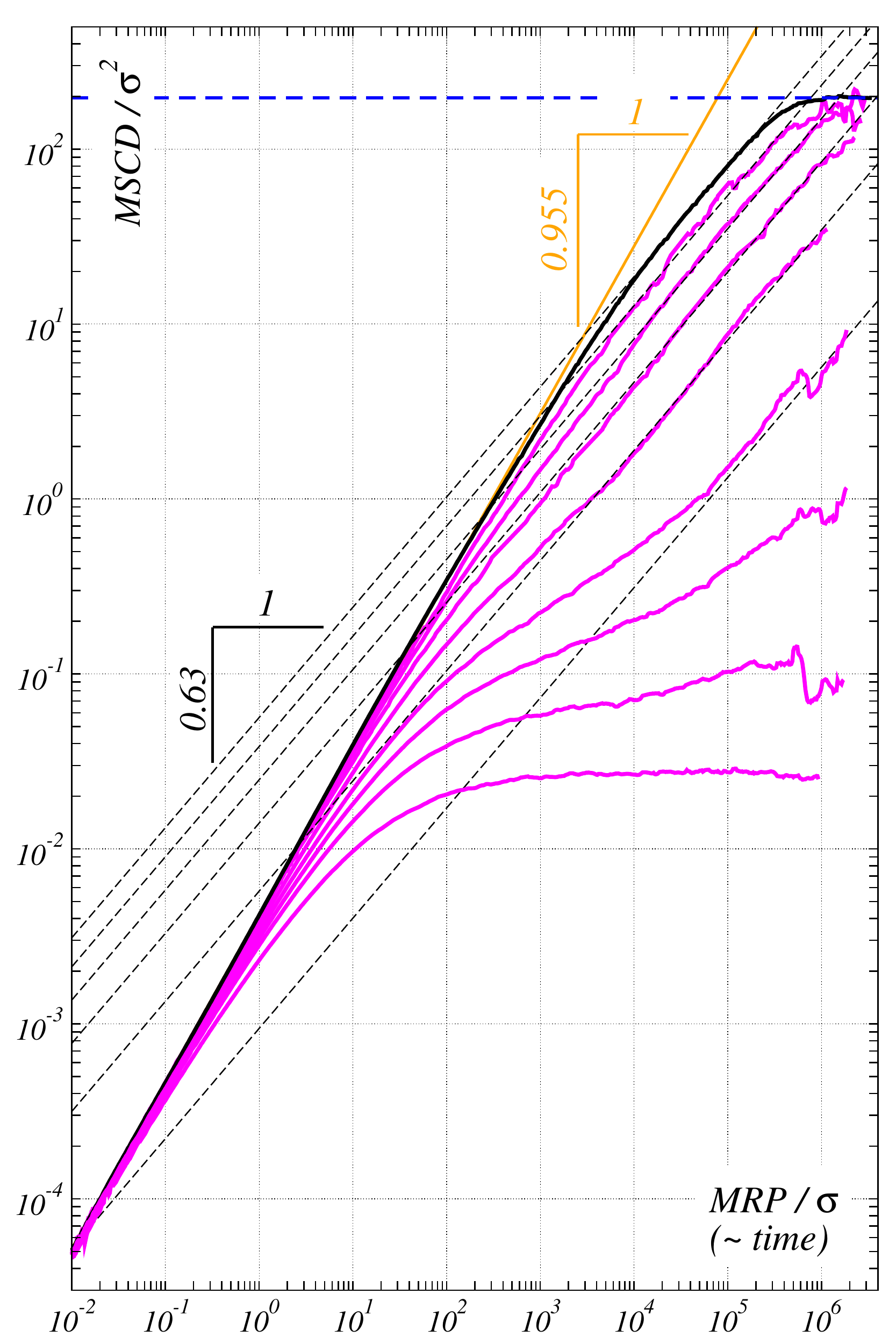}
\vspace{-1em}
\caption{Reduced, dimensionless particle mean squared chemical displacement
($MSCD / \sigma^2 = \left\langle\right. \delta l^2 \left.\right\rangle / \sigma^2$)
as a function of the reduced, dimensionless mean rectified path ($MRP / \sigma$) which is proportional to time.
Uppermost, black curve: Single particle dynamic Monte Carlo ($\phi=0$).
Pink curves: Dynamic Monte Carlo of fluids at packing fractions $\phi = 0.1, ~0.2, ~0.3, ~0.4, ~0.5, ~0.55, ~0.6$ and $0.65$
(from top to bottom).
Solid orange and dashed black lines: Fitted short-time and long-time power laws, respectively.
All dashed black lines are for the same exponent $0.63$.
Blue dashed line: $MSCD = 196 ~\sigma^2$ for particles that sample the length scale of the periodic simulation box.}
\label{fig:MSCD_vs_MRP}
\end{figure}

As shown in Fig. \ref{fig:MSCD_vs_MRP}, the simulations reveal anomalous diffusion at all time scales, with an MSCD that increases as
$\left\langle\right. \delta l^2 (t) \left.\right\rangle \propto t^{\alpha(t)}$, where $\alpha(t \to 0) = 0.96(1)$ and $\alpha(t \to \infty) = 0.63(1)$
are the coefficients of anomalous short- and long-time self-diffusion, respectively. Both exponents are found for an ideal gas of non-interacting,
extended (\textit{i.e.} non point-like) fractal particles in the limit of vanishing packing fraction ($\phi \to 0$) and for concentrated systems alike.
The uncommon presence of two different exponents for the short- and long-time diffusion of \emph{non-interacting} particles can be attributed to the
breaking of configuration-space scale invariance by the particle length scale $\sigma$, and this conjecture is supported by the observation that the
crossover between the two different power laws occurs at an MSCD which is of the order of $\sigma^2$ (\textit{c.f.} the uppermost black curve in Fig. \ref{fig:MSCD_vs_MRP}).
For increasing particle packing fractions (pink curves in Fig. \ref{fig:MSCD_vs_MRP}), the short-time anomalous diffusion and the long-time anomalous diffusion regimes
get separated by an intermediate sub-diffusive regime during which the particles sample and break out of their nearest neighbor cages. For high packing fractions of
$\phi \geq 60\%$ (lowermost two pink curves in Fig. \ref{fig:MSCD_vs_MRP}), particles are localized within a small MSCD during the entire simulation which
covers eight decades in time. The observed value of the MSCD for localized particles is of the order of $(0.1 \sigma)^2$, in accordance with the 
classical Lindemann melting criterion. It is worthwhile to notice the ratio $\alpha(t \to \infty) / \alpha(t \to 0) = 2/3$ within the limited precision
of the present measurements. In spite of the new, improved dynamic MC method, the simulations remain numerically expensive: Every curve in  Fig. \ref{fig:MSCD_vs_MRP} represents
an ensemble average over $32$ statistically independent system trajectories, and every such trajectory requires approximately $2$ weeks of simulation time
on a single CPU core.

\begin{figure}
\includegraphics[width=.99\columnwidth]{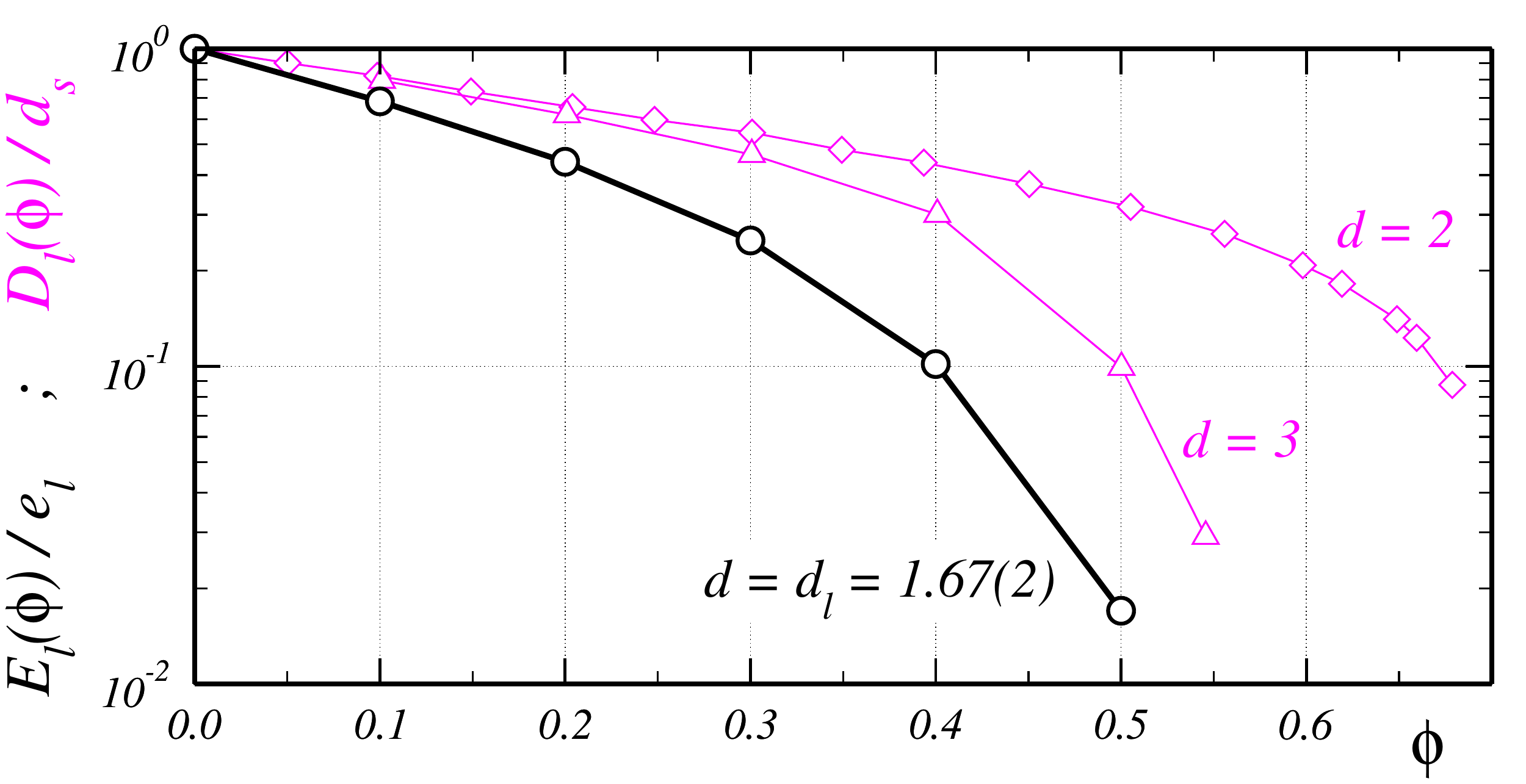}
\vspace{-1em}
\caption{Black: Reduced, dimensionless coefficient of anomalous long-time self-diffusion for the fractal particles studied in this work.
Pink: Reduced, dimensionless coefficients of normal long-time self diffusion for hard disks in $d=2$ \cite{Thorneywork2015} (diamonds) and
hard spheres in $d=3$ \cite{Cichocki1990, Lowen1993, Sanz2010} (triangles).
All results are plotted as functions of the dimensionless particle packing fractions $\phi$. 
Lines are guides to the eye.}
\label{fig:Transport_coeff_vs_phi}
\end{figure}

The coefficient $E_l(\phi)$ of anomalous long-time self diffusion can be defined by the equation
$\left\langle\right. \delta l^2 (t \to \infty) \left.\right\rangle =  E_l(\phi) ~ t^{\alpha(t \to \infty)}$ and the corresponding
value for the ideal gas is denoted here as $e_l = E_l(\phi = 0)$. Figure \ref{fig:Transport_coeff_vs_phi} features
the reduced, dimensionless coefficient $E_l(\phi) / e_l$ for packing fractions $\phi \leq 0.5$, along with the dimensionless coefficients
for normal long-time self-diffusion of monodisperse hard disks in $d=2$ \cite{Thorneywork2015} and monodisperse hard spheres in $d=3$
\cite{Cichocki1990, Lowen1993, Sanz2010}. Comparing coefficients of anomalous and normal diffusion in one plot, there is no reason
to expect monotonic ordering of the coefficients with respect to dimension. In fact, the dimensionless coefficient of anomalous long-time
self diffusion in $d = d_l = 1.67(2)$ attains numerical values that are smaller than those of the dimensionless coefficient for the normal
long-time self-diffusion in $d=3$. The latter coefficient, in turn, is smaller than the coefficient for $d=2$. 

In conclusion, the anomalous diffusive dynamics of particles in a $1.67(2)$-dimensional equilibrium fluid reveals itself to be 
rich in peculiarities: Short- and long-time anomalous self diffusion are characterized by two exponents of the MSCD as a function of
time, and the exponents exhibit a ratio of $3 : 2$ within an accuracy level of $1\%$. The numerical values of the exponents, $0.96(1)$ and $0.63(1)$,
have not been reported anywhere else to my best knowledge. Furthermore, the distinction between short- and long-time self diffusion
prevails even in the case of an ideal gas of non-interacting particles, which exhibit the same two (short- and long-time) exponents.

A theory for fractal fluid diffusion is still at demand, and a promising approach might be the application of fractional calculus
\cite{Metzler2000, Balankin2017} to a Smoluchowski equation for interacting Brownian particles \cite{Dhont1996, Lionberger2000}
in non-integer dimensional space. Another viable approach could be the analytical continuation of equilibrium and non-equilibrium self-consistent
generalized Langevin equation theory \cite{Olais-Govea2015} with respect to dimension, which might offer a pathway to studying fractal
nonequilibrium state transitions into gels and glasses. Whether or not mode-coupling theory \cite{Krakoviack2005, Schnyder2011} can be generalized
to the case of fractal particles in a likewise fractal configuration space is presently unclear.  
Future work will have to probe the existence of equilibrium phase transitions such as a liquid-crystal transition.
Although symmetry breaking transitions are ruled out in dimensions $d \leq 2$ by the Mermin-Wagner theorem, the possibility of infinite order
phase transitions like the ones observed in $d=2$ \cite{Kosterlitz1973} cannot be ruled out.

\section*{Acknowledgement}
It is my pleasure to thank
Simon K. Schnyder,
C\'{e}sar A. B\'{a}ez,
Magdaleno Medina-Noyola,
Francisco Sastre Carmona,
Hartmut L\"{o}wen,
and John F. Brady and the members of his team
for our many discussions. 
Part of this work was supported by a fellowship within the Postdoc-Program of the German Academic Exchange Service (DAAD).
I gratefully acknowledge the computing time granted by LANCAD and CONACYT on the supercomputer miztli at DGTIC UNAM.


\begin{thebibliography}{99}

\bibitem{Neser1997}
S.~Neser, C.~Bechinger, P.~Leiderer, and T.~Palberg.
\newblock {\em Phys. Rev. Lett.}, 79:2348--2351, 1997.

\bibitem{Steward2014}
M.~C. Stewart and R.~Evans.
\newblock {\em J. Chem. Phys.}, 140:134704, 2014.

\bibitem{Zahn2000}
K.~Zahn and G.~Maret.
\newblock {\em Phys. Rev. Lett.}, 85:3656--3659, 2000.

\bibitem{Deutschlaender2013}
S.~Deutschl\"{a}nder, T.~Horn, H.~L\"{o}wen, G.~Maret, and P.~Keim.
\newblock {\em Phys. Rev. Lett.}, 111:098301, 2013.

\bibitem{Hahn1996}
K.~Hahn, J.~K\"{a}rger, and V.~Kukla.
\newblock {\em Phys. Rev. Lett.}, 76:2762--2765, 1996.

\bibitem{Wei2000}
Q.-H. Wei, C.~Bechinger, and P.~Leiderer.
\newblock {\em Science}, 287:625--627, 2000.

\bibitem{Lutz2004}
C.~Lutz, M.~Kollmann, and C.~Bechinger.
\newblock {\em Phys. Rev. Lett.}, 93:026001, 2004.

\bibitem{benAvraham_Havlin2000}
D.~ben Avraham and S.~Havlin.
\newblock {\em Diffusion and Reactions in Fractals and Disordered Systems}.
\newblock Cambridge University Press, Cambridge CB2 2RU, UK, 2000.

\bibitem{Levitt1973}
D.~G. Levitt.
\newblock {\em Phys. Rev. A}, 8:3050, 1973.

\bibitem{Karger1992}
J.~K\"{a}rger.
\newblock {\em Phys. Rev. A}, 45:4173--4174, 1992.

\bibitem{Pape1999}
H.~Pape, C.~Clauser, and J.~Iffland.
\newblock {\em Geophysics}, 64, 1999.

\bibitem{Hoefling2013}
F.~H\"{o}fling and T.~Franosch.
\newblock {\em Rep. Prog. Phys.}, 76:046602, 2013.

\bibitem{Zhou2012}
Z.~Zhou, J.~Yang, Y.~Deng, and R.~M. Ziff.
\newblock {\em Phys. Rev. E}, 86:061101, 2012.

\bibitem{Krakoviack2005}
V.~Krakoviack.
\newblock {\em Phys. Rev. Lett.}, 94:065703, 2005.

\bibitem{Skinner2013}
T.~O.~E. Skinner, S.~K. Schnyder, D.~G. A.~L. Aarts, J.~Horbach, and R.~P.~A.
  Dullens.
\newblock {\em Phys. Rev. Lett.}, 111:128301, 2013.

\bibitem{Heinen2015}
M.~Heinen, S.~K. Schnyder, J.~F. Brady, and H.~L\"{o}wen.
\newblock {\em Phys. Rev. Lett.}, 115:097801, 2015.

\bibitem{Percus1958}
J.~K. Percus and G.~J. Yevick.
\newblock {\em {Phys. Rev.}}, {110}:{1--13}, {1958}.

\bibitem{Heinen2014}
M.~Heinen, E.~Allahyarov, and H.~L\"{o}wen.
\newblock {\em J. Comput. Chem.}, 35:275--289, 2014.

\bibitem{Santos2016}
A.~Santos and M.~Lopez~de Haro.
\newblock {\em Phys. Rev. E}, 93:062126, 2016.

\bibitem{Dijkstra1959}
E.~W. Dijkstra.
\newblock {\em Numerische Mathematik}, 1:269--271, 1959.

\bibitem{Cichocki1990}
B.~Cichocki and K.~Hinsen.
\newblock {\em Physica A}, 166:473--491, 1990.

\bibitem{Sanz2010}
E.~Sanz and D.~Marenduzzo.
\newblock {\em J. Chem. Phys.}, 132:194102, 2010.

\bibitem{Thorneywork2015}
A.~L. Thorneywork, R.~E. Rozas, R.~P.~A. Dullens, and J.~Horbach.
\newblock {\em Phys. Rev. Lett.}, 115:268301, 2015.

\bibitem{Lowen1993}
H.~L\"{o}wen and G.~Szamel.
\newblock {\em J. Phys.-Condes. Matter}, 5:2295--2306, 1993.

\bibitem{Metzler2000}
R.~Metzler and J.~Klafter.
\newblock {\em Phys. Rep.}, 339:1--77, 2000.

\bibitem{Balankin2017}
A.~S. Balankin, B.~Mena, O.~Susarrey, and D.~Samayoa.
\newblock {\em Phys. Lett. A}, 381:623--628, 2017.

\bibitem{Dhont1996}
J.~K.~G. Dhont.
\newblock {\em An Introduction to Dynamics of Colloids}.
\newblock Elsevier, Amsterdam, 1996.

\bibitem{Lionberger2000}
R.~A. Lionberger and W.~B. Russel.
\newblock Microscopic theories of the rheology of stable colloidal dispersions.
\newblock In {\em Advances in Chemical Physics}, volume 111, pages 399--474.
  John Wiley \& Sons Inc., NY 10016 USA, 2000.

\bibitem{Olais-Govea2015}
J.~M. Olais-Govea, L.~L\'{o}pez-Flores, and M.~Medina-Noyola.
\newblock {\em J. Chem. Phys.}, 143:174505, 2015.

\bibitem{Schnyder2011}
S.~K. Schnyder, F.~H\"{o}fling, T.~Franosch, and T.~Voigtmann.
\newblock {\em J. Phys.: Condens. Matter}, 23:234121, 2011.

\bibitem{Kosterlitz1973}
J.~M. Kosterlitz and D.~J. Thouless.
\newblock {\em J. Phys. C.}, 6:1181, 1973.

\end{thebibliography}

\end{document}